\documentclass{pic2012}
\def\runauthor{E. Polycarpo}
\def\shorttitle{Charm mixing and CPV}
\usepackage{amsmath}
\usepackage{url}
\begin{document}

\title{$D^0$ MIXING AND CP VIOLATION IN $D$ DECAYS
}

\author{\'Erica Polycarpo}

\address{Universidade Federal do Rio de Janeiro\\
CP 68528, 20511-010, Rio de Janeiro, RJ, Brasil\\
E-mail: poly@if.ufrj.br}

\maketitle

\abstracts{We present a brief review of CPV and mixing measurements in the charm sector, with emphasys in results published or presented since the previous edition of the Physics in Collision Symposia.
}

\section{Introduction} 

\begin{figure}[ph]
\begin{center}
\begin{tabular}{cc}
\includegraphics[width=0.48\textwidth]{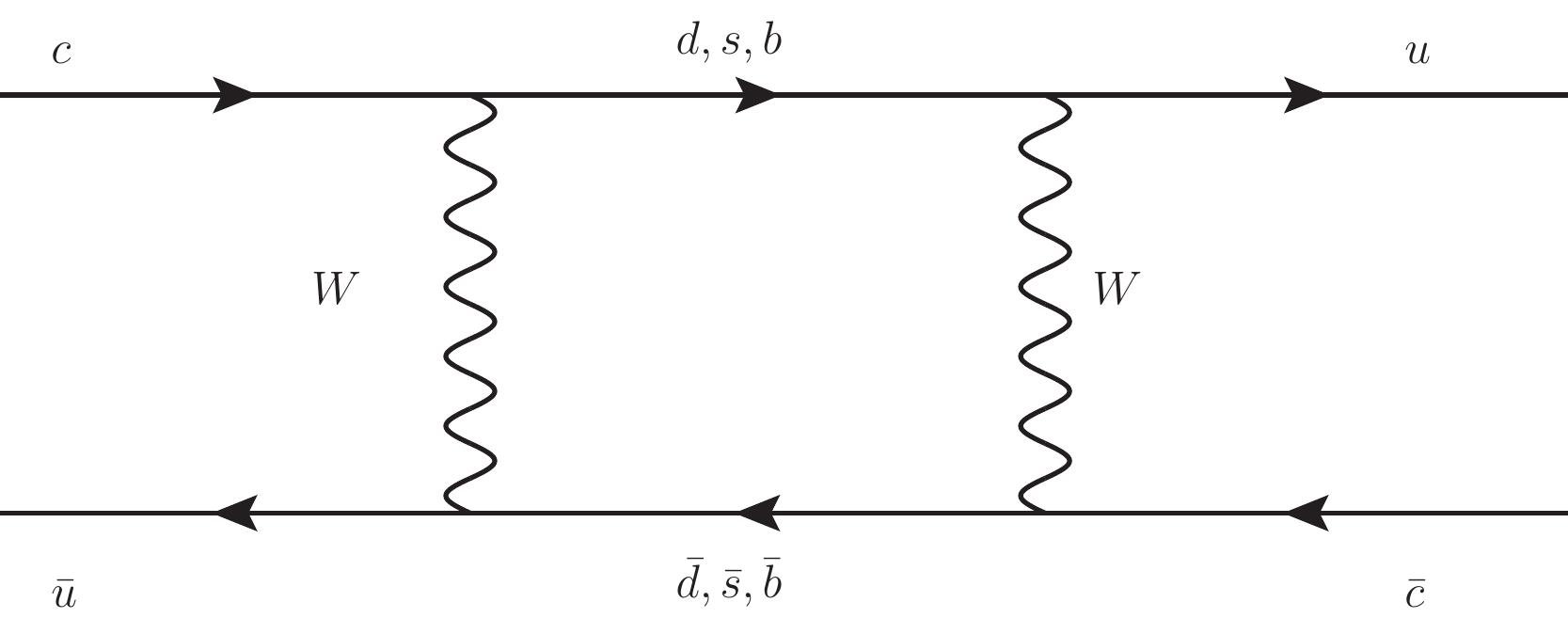}&
\includegraphics[width=0.48\textwidth]{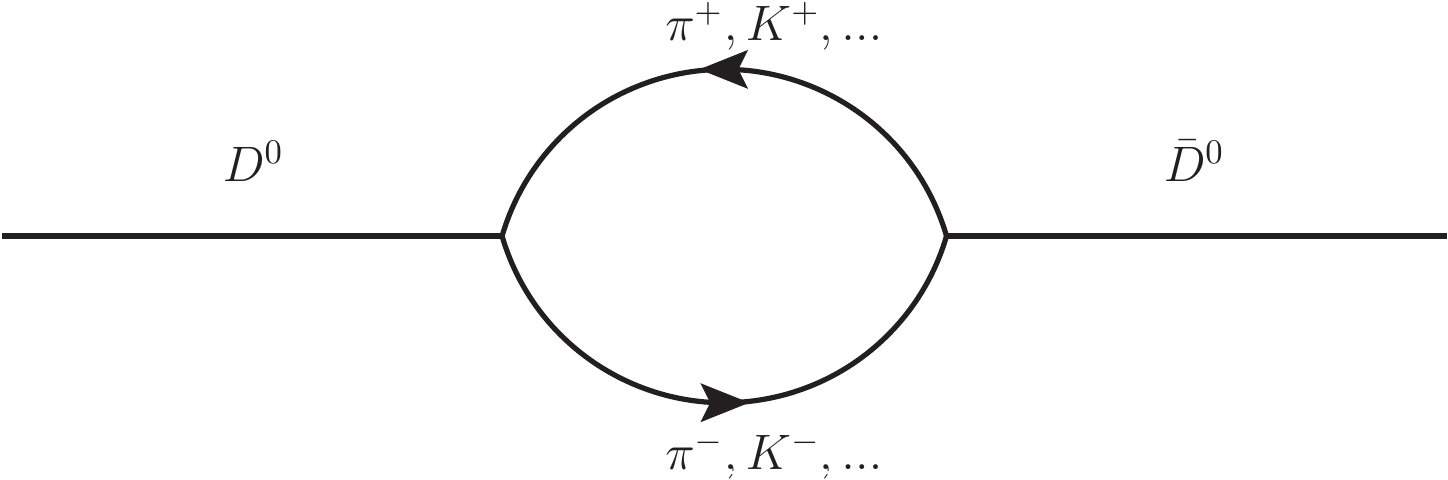}\\
\end{tabular}
\caption[*]{Short and long distance contributions to $D^0-\bar{D}^0$ mixing.}\label{fig:mixingdiagrams} 
\end{center}
\end{figure}

Neutral mesons mixing occur in nature due to the fact that the mass eigenstates are not weak interaction eigenstates. 
An initially pure beam of $D^0$ or $\bar{D}^0$ mesons evolves in time to a mixture of $D^0$ and $\bar{D}^0$ states with rates determined by the solution of the Schr\"oedinger equation
\begin{equation}
i\frac{d}{dt}
\begin{pmatrix}
D^0  \\
\bar{D}^0\\
\end{pmatrix}(t)={\bf H_{eff}} 
\begin{pmatrix}
D^0 \\
\bar{D}^0\\
\end{pmatrix}(t),
\label{eq:schrodinger}
\end{equation}
where the diagonal elements of the effective Hamiltonian $\bf H_{eff}\equiv M - \frac{i}{2}\Gamma$ and of the hermitian matrices $M$ and $\Gamma$ are related to the flavour conserving $D^0(\bar{D}^0)\rightarrow D^0(\bar{D}^0)$ transitions, while the non-diagonal ones are related to the flavour changing $D^0\leftrightarrow \bar{D}^0$ transitions. The solution of Eq.\ref{eq:schrodinger} are the eigenstates 
$|D_{1,2}\rangle= p|D^0\rangle \pm q|\bar{D}^0\rangle$,
with complex coefficients $p$ e $q$ satisfying $\sqrt{|p|^2+|q|^2}=1$. The eigenstates have well defined masses and widths $m_1,\Gamma_1$ and $m_2,\Gamma_2$. The parameters which govern the mixing are $x\equiv \frac{m1-m2}{\Gamma}$ and $y\equiv \frac{\Gamma_1-\Gamma_2}{2\Gamma}$, where $\Gamma\equiv \frac{\Gamma_1+\Gamma_2}{2}$ is the mean decay width. For charm mesons, both $x$ and $y$ are small. There are essentially two approaches used to predict their values in the Standard Model (SM), none of them leading to fully reliable results (see, for example,~\cite{lenz.ichep2010,petrov.2002.mix,gersabeck.2011} and references therein). The transitions leading to the $D^0-\bar{D}^0$ oscillations are those represented in Fig.~\ref{fig:mixingdiagrams}. Short distance contributions are illustrated by the box diagram on the left and long distance transitions, involving intermediate states accessible by both $D^0$ and $\bar{D}^0$, on the right. As seen from these diagrams, charm mixing gives us access to transitions involving intermediate states with down-type quarks. In fact, this is the only system which offers this probe, since the top quark is too heavy to hadronize and the $\pi^0$ is its own anti-particle. 

While $x$ receives contributions from both the short distance and long distance processes, $y$ only receives contributions from the long distance transitions. Since $x$ is the only one to receive contributions from virtual intermediate states, new physics introducing off-shell particles in the box diagrams would only affect $x$ and therefore it is a sort of consensus that an unambiguous sign of NP would be given by a $x$ value well above $y$. The first evidences of mixing were obtained in 2007 by BaBar~\cite{babar.mix}, Belle~\cite{belle.mix} and CDF~\cite{cdf.mix}. At present, both $x$ and $y$ are constrained by several measurements at the values $x=(0.65^{+.018}_{-0.19})$\% and $y=(0.73\pm 0.12)$\% and the no mixing hypothesis is excluded with a statistical significance above 10 standard deviations ($\sigma$)~\cite{neri.charm2012}, under the assumption of no CP violation. Although theoretical predictions are not precise, the experimental values of the mixing parameters are now consistent with SM expectations~\cite{mixing.expectations}, even if still considered at the upper level of possible values~\cite{lhcb.implications}.

The CP symmetry can be violated in the mixing, if $|q|\neq |p|$; in the decay, if the magnitudes of the instantaneous decay amplitudes ${\cal A}_f=\langle f|{\cal H}|D^0 \rangle$ and $\bar{\cal A}_{\bar{f}}=\langle \bar{f}|{\cal H}|\bar{D}^0 \rangle$ for CP conjugate processes are not the same; and, for decays reachable by both the $D^0$ and $\bar{D}^0$, in the interference between mixing and decay. This kind of mixing induced CPV can occur even in the absence of CPV in the mixing and in the decay, if there is a non null phase difference $\phi$ between the mixing and the decay amplitudes. 

In the SM, CP violation (CPV) is suppressed by the small values of the CKM parameters $V_{ub}$ and $V_{cb}$. In the mixing, it is expected to be at the level of 10$^{-4}$~\cite{bigi.2011.cpv}, independently of the final state. Direct CPV can only occur in singly Cabibbo suppressed (SCS) decays, since their amplitudes can receive contributions from processes with virtual b-quarks, involving the complex elements of the CKM matrix. Existing predictions for SCS modes in the SM were, at least until recently, of the order of $10^{-3}$ or less. The interest in CPV in charm decays is, therefore, related to searches for new physics effects in a SM suppressed environment. Experimentally, the only existing evidence of CPV in the charm sector is the measurement of the difference between the time integrated asymmetries of the decays $D^0\to K^+K^-$ and $D^0\to\pi^+\pi^-$~\cite{lhcb.dacp}, that will be further discussed in this document. 

A very good review on CPV and mixing on the charm sector is given in~\cite{gersabeck.2011}. Here we report on some recent measurements performed by BaBar, Belle, CDF and LHCb since the previous edition of the Physics in Collision conference series. 

\section{Time dependent measurements of $D^0-\bar{D}^0$ mixing and CP violation}

There are several ways to access the mixing parameters $x$ and $y$ experimentally, using decays to various final states. Each method provide different sensitivities to these parameters and, by combining these measurements, we can improve our knowledge about them. The HFAG group~\cite{hfag.site} provides world averages and best fit values to these parameters using results provided by several experiments at different machines around the world. A good overview of the most up to date averages are given in~\cite{neri.charm2012}. 


One of the most sensitive ways of determining $x$ and $y$ is via a time dependent amplitude analysis of the decays $D^0\to K_s \pi^+\pi^-$ and $D^0\to K_sK^+K^-$. The two final states are CP eigenstates reachable by both the $D^0$ and $\bar{D}^0$. The time dependent rate for these decays is proportional to terms arising from the interference between the direct decay and the decay via mixing:
\begin{equation}
{\mathcal R}(t)\propto { |A_1|^2 e^{-y t}+|A_2|^2 e^{-y t}+2Re[A_1 A^*_2]\cos(xt)+2Im[A_1 A^*_2]\sin(xt) }  
\end{equation}
where $A_{1,2}=\frac{{\cal A}\pm\bar{\cal A}}{2}$ and ${\cal A}$ and $\bar{\cal A}$ are the instantaneous decay amplitudes for $D^0$ and $\bar{D}^0$, written as a function of the 2-body invariant masses $M^+$ ($K^0_SK^+$) and $M^-$ ($K^0_sK^-$). In order to determine the flavour of the meson at production, it is required to form a vertex with a low momentum pion $\pi_s$, consistent with the decay vertex of a $D^{*+}\to\pi^+_sD^0$ or $D^{*-}\to\pi^-_s\bar{D}^0$. The charge of the $\pi_s$ identifies the $D^0$ flavour. This analysis was first performed by CLEO~\cite{cleo.kspp} and subsequently by Belle~\cite{belle.kspp} and BaBar~\cite{babar.kskkpp}, with much larger samples. An update of the $D^0\to K_s \pi^+\pi^-$ analysis by Belle was presented this summer, using their full data set of 920 fb$^{-1}$ and an improved tracking, which provides a signal yield of 1.23 M events with a purity of 95.6\%~\cite{belle.kspipi.ichep}. They parametrize the amplitudes ${\cal A}$ and $\bar{\cal A}$ as a sum of 12 quasi-two-body amplitudes describing the P and D waves and two components to describe the $K\pi$ (LASS model) and the $\pi\pi$ (K-matrix model) s-waves. From a maximum likelihood fit to the data in the signal region, the mixing parameters $x$ and $y$, the average $D^0$-meson lifetime $\tau_{D^0}$ and the magnitudes and phases of the resonances are directly extracted. 
The preliminary results shown are $x=(0.56\pm 0.19^{+0.03+0.06}_{0.09-0.09})$\%  and $y=(0.30\pm 0.15^{+0.04+0.03}_{-0.05-0.06})$\%, where the first uncertainty is statistical, the second is due to experimental systematic effects and the third systematical due to the model used to fit the Dalitz plot. At the time of the conference, this was the most precise single measurement of the mixing parameters.

Another measurement which is sensitive to mixing is the deviation from unity of the ratio of the effective lifetimes measured in decays to the CP eigenstates $K^-K^+$ and $\pi^-\pi^+$ with respect to the CP mixed state $K ^-\pi^+$. Since mixing is very slow, the decay time distributions are, to a good approximation, simply exponential. The fits to the decay time distributions of the different samples use a model where signal events are described by a single exponential and provide the effective lifetimes used to build the observable
\begin{equation}
y_{CP}\equiv \frac{\tau(D^0\to h^-h^+)}{\tau(D^0\to K^-\pi^+)}-1\approx \frac{1}{2}\left[\left( \begin{vmatrix}\frac{q}{p}\end{vmatrix}+\begin{vmatrix}\frac{p}{q}\end{vmatrix}\right)y\cos\phi- \left(\begin{vmatrix}\frac{q}{p}\end{vmatrix}-\begin{vmatrix}\frac{p}{q}\end{vmatrix}\right)x\sin\phi\right]
\end{equation}
where $h^-h^+$ corresponds to $\pi^-\pi^+$ or $K^-K^+$, the phase $\phi$ is considered universal and a contribution from CPV in the decay is neglected. In the absence of CPV, $|q/p|=1$ and $\phi=0$, hence $y_{CP}=y$. This measurement was previously performed by several collaborations and provided the first evidence for mixing obtained by the BELLE experiment in 2007~\cite{belle.mix}. In April 2011, both BaBar and Belle presented new results using their full data samples, with integrated luminosities of 468$\,$fb$^{-1}$~\cite{babar.ycp} and 976$\,$fb$^{-1}$~\cite{belle.ycp}, respectively. LHCb also published a first result for $y_{CP}$ using~29$\,$pb$^{-1}$ of data collected in 2010~\cite{lhcb.ycp}. In addition to the $D^{*\pm}$ tagged samples, BaBar used also untagged samples of $D^0(\bar{D}^0)\to K^\pm\pi^\mp$ and $D^0(\bar{D}^0)\to K ^-K^+$ decays. These results are presented in Table~\ref{tab:ycp.ag}, together with the size of the samples used in each measurement. 

The effective lifetimes obtained by fitting the decay time distributions of the tagged samples separately for the CP conjugate decays provide a second quantity, $A_{\Gamma}$, sensitive to CPV:
\begin{equation}
\small
A_\Gamma (h^-h^+)\equiv \frac{\tau(\bar{D}^0\to h^-h^+)- \tau(D^0\to h^-h^+)}{\tau(\bar{D}^0\to h^-h^+)+\tau(D^0\to h^-h^+)}\approx \frac{1}{2}\left[\left( \begin{vmatrix}\frac{q}{p}\end{vmatrix}-\begin{vmatrix}\frac{p}{q}\end{vmatrix}\right)y\cos\phi- \left( \begin{vmatrix}\frac{q}{p}\end{vmatrix}+\begin{vmatrix}\frac{p}{q}\end{vmatrix}\right)x\sin\phi\right],
\end{equation}  
considering as valid the same approximations used for $y_{CP}$. 

The three results for $y_{CP}$ and $A_\Gamma$ are consistent with each other and with previous measurements and also compatible with the hypothesis of no CPV. They bring down the world average for $y_{CP}$ from $(1.064\pm0.209)$\% to $(0.866\pm0.155)$\%~\cite{neri.charm2012}, relaxing the previously existing tendency  for $y_{CP}>y$~\cite{gersabeck.2011}. The new world average for $A_\Gamma$ is $(-0.022\pm 0.161)$\%~\cite{neri.charm2012}.  

\begin{table}[h]
\begin{center}
\caption{Recent experimental results for the quantities $y_{CP}$ and $A_\Gamma$. The $h^-h^+$ yield is given for $K^-K^+$ and inside the parentheses for $\pi^-\pi^+$. First uncertainty is statistical, second is systematical. BaBar measures, instead of $A_\Gamma$, $\Delta Y=(1+y_{CP})A_\Gamma$.}\protect\label{tab:ycp.ag}
\small
\begin{tabular}{ccccc}\hline
Experiment &  $K^-\pi^+$ & $h^-h^+$ & $y_{CP}$ (\%) & $A_\Gamma$ ($\Delta Y$) (\%)\\
          &  yield ($\times 10^6$)& yield ($\times 10^3$)& & \\\hline
LHCb       &   0.225 & 30  & $5.5\pm 6.3\pm 4.1$   & $-5.9\pm 5.9\pm 2.1$ \\ 
BaBar      &   7.312 & 633 (65) & $0.72\pm 0.18\pm0.12$ & $0.09\pm 0.26\pm 0.09$ \\ 
Belle      &   2.61  & 242 (114)& $1.11\pm 0.22\pm 0.11$ & $-0.03\pm 0.20\pm 0.08$ \\\hline 
\end{tabular}
\end{center}
\end{table}

\section{Time integrated CP asymmetries}

\subsection{Two-body decays}

One of the most discussed results from the LHC in the past year was the first evidence of CPV in the charm sector, provided by the difference of the time integrated CP asymmetries measured in the decays of the $D^0$ meson to $K^-K^+$ and $\pi^-\pi^+$, presented by LHCb~\cite{lhcb.dacp} and supported by new results by CDF~\cite{cdf.dacp} and Belle~\cite{belle.dacp}. To first order of approximation, the raw asymmetry can be written as a sum of small terms
\begin{equation}
A_{raw}(f)=A_{CP}(f)+ A_{det}(f) + A_{det}(\pi_s) + A_{prod}(D^{*+}),
\end{equation}
where $A_{CP}(f)$ is the physical CP asymmetry of interest, $A_{det}(f)$ is the asymmetry in the detection efficiency for the decay to final state $f$, $A_{det}(\pi_s)$ is the reconstruction efficiency for the slow pion from the $D^{*+}\to \pi^+_s D^0$ tagging chain and $A_{prod}(D^{*+})$ is a physical asymmetry for $D^{*+}$ production. This asymmetry vanishes in CP conserving strong collisions, as the $p\bar{p}$ collisions at the Tevatron, if the events are reconstructed symmetrically in the pseudo-rapidity $\eta$. However, it is not expected to be null in the $pp$ collisions at the LHC. In $e^+e^-$ colliders, there is a forward-backward asymmetry in the production of charm mesons arising from $\gamma Z^0$ interference and higher order QED effects~\cite{bfactories.afb}.  

When dealing with two-body decays of a spin 0 particle to self-conjugate final states $f$, the detection efficiency $A_{det}(f)$ is null. Since the remaining terms, apart from $A_{CP}(f)$, are independent of the final state, they cancel out in the difference of asymmetries $\Delta A_{raw}=A_{raw}(K^+K^-)-A_{raw}(\pi^+\pi^-)$, which thus becomes a robust measurement of the difference between the physical CP asymmetries: 
\begin{equation}
\Delta A_{CP}\equiv A_{CP}(K^+K^-)-A_{CP}(\pi^+\pi^-)= \Delta A_{raw}. 
\end{equation}
Experimentally, the raw asymmetries are obtained from 
\begin{equation}
A_{raw}(f)\equiv \frac{N(D^0\to f)-N(\bar{D}^0\to f)}{N(D^0\to f)+N(\bar{D}^0\to f)}  
\end{equation}
where $N(X)$ corresponds to the number of reconstructed decays $X$ after background subtraction. 

     LHCb performed this measurement using 0.62 fb$^{-1}$ of data collected in 2011. Maximum likelihood fits to the $\delta m=m(\pi_s h^-h^+)-m(h^-h^+)$ spectra provide the number of signal events used to calculate the raw asymmetries in different kinematic bins, in order to guarantee that the detection and production asymmetries are cancelled. The kinematic bins are defined in intervals of the $D^{*\pm}$ candidate transverse momentum $p_T$ and $\eta$ and the final result, shown in Table~\ref{tab:dacp}, is obtained from a weighted average over this kinematic space. This result deviates from zero with a significance of 3.5 $\sigma$.
       
    CDF also performed a new $\Delta A_{CP}$ measurement, using the full data sample collected at Tevatron Run II, which corresponds to 9.7 fb$^{-1}$~\cite{cdf.dacp}. Compared to the previously published measurement~\cite{cdf.acp.2011}, their new result relies on a larger data sample and an optimized selection procedure. In order to take into account differences in the instrumental asymmetries due to phase space variations among the two final states, a reweighting of the samples is performed, so as to equalize the $\pi_s$ impact parameter, $p_T$ and $\eta$ distributions. A simultaneous $\chi^2$ fit of the resulting $D^0\pi^+$ and $\bar{D}^0\pi^-$ mass distributions provides the number of signal events used in the raw asymmetries. The result, also shown in Table~\ref{tab:dacp}, is 2.7$\sigma$ away from zero, with a precision level very close to the obtained by LHCb. The individual asymmetries, measured with a sample of 6.0 fb$^{-1}$, are $A_{CP}(K^-K^+)=-0.24\pm 0.22\pm 0.09$\% and $A_{CP}(\pi^-\pi^+)=(+0.22\pm 0.24\pm 0.11)$\%~\cite{cdf.acp.2011}. 

   In July 2012, Belle presented a preliminary update of the individual asymmetries $A_{CP}(K^-K^+)$ and $A_{CP}(\pi^-\pi^+)$, reporting also a $\Delta A_{CP}$ measurement, using a data sample corresponding to 976 fb$^{-1}$. Instrumental asymmetries are obtained using control samples of CP conserving decays and the production asymmetry is eliminated by making the measurement in bins of $cos\theta^*$, where $\theta^*$ is the polar angle of the $D^{*+}$ production in the center of mass system. The results are $A_{CP}(K^-K^+)=-0.3\pm 0.21\pm 0.09$\% and $A_{CP}(\pi^-\pi^+)=(+0.55\pm 0.36\pm 0.09)$\%. The $\Delta A_{CP}$ value, given in Table~\ref{tab:dacp}, deviates from zero with a 2.1$\sigma$ significance. 

\begin{table}[h]
\begin{center}
\caption{Recent experimental results for $\Delta A_{CP}$. The last column shows the difference between the average proper time of the $K^-K ^+$ and $\pi^-\pi^+$ samples used to extract $\Delta A_{CP}$.}\protect\label{tab:dacp}
\small
\begin{tabular}{ccccc}\hline
Experiment &  $K^-K^+$  & $\pi^-\pi^+$  & $\delta A_{CP}$ & $\Delta \langle t\rangle$\\
           &  yield ($\times 10^6$) & yield ($\times 10^6$) & & \\\hline 
LHCb       &  1.44  & 0.38 & ($−0.82\pm 0.21\pm 0.11$)\% &  $\sim 0.098 \tau_{D^0}$ \\\ 
CDF        &  0.55  & 1.2 &  ($-0.62\pm0.21\pm0.10$)\%  & $\sim 0.26 \tau_{D^0}$\\
Belle      &  0.282 & 0.123 &  ($-0.87\pm0.41\pm0.06$)\% &  0 \\\hline 
\end{tabular}
\end{center}
\end{table}  

 The asymmetry difference can be written, to first order of approximation, in terms of the direct and indirect components as~\cite{cdf.acp.2011} 
\begin{equation}
\Delta A_{CP}\approx \Delta a^{dir}_{CP} + \frac{\Delta \langle t\rangle}{\tau_{D^0}} a^{ind} ,
\label{eq:dacp}
\end{equation}
where $\langle t\rangle$ is the average reconstructed decay time, $a^{dir}_{CP}(f)=(|{\cal A}_f|^2- |\bar{\cal A}_f|^2)/(|{\cal A}_f|^2+|\bar{\cal A}_f|^2)$ and $a^{ind}_{CP}=-\frac{1}{2}\left[\left( \begin{vmatrix}\frac{q}{p}\end{vmatrix}-\begin{vmatrix}\frac{p}{q}\end{vmatrix}\right)y\cos\phi- \left( \begin{vmatrix}\frac{q}{p}\end{vmatrix}+\begin{vmatrix}\frac{p}{q}\end{vmatrix}\right)x\sin\phi\right]$ and the phase $\phi$ is assumed to be universal. From this equation and the $\Delta\langle t\rangle/\tau_{D^0}$ values given in Tab.~\ref{tab:dacp} for the different experiments, it becomes clear that $\Delta A_{CP}$ is mostly a measurement of direct CPV.

\begin{figure}
\begin{center}
\includegraphics[width=0.5\textwidth]{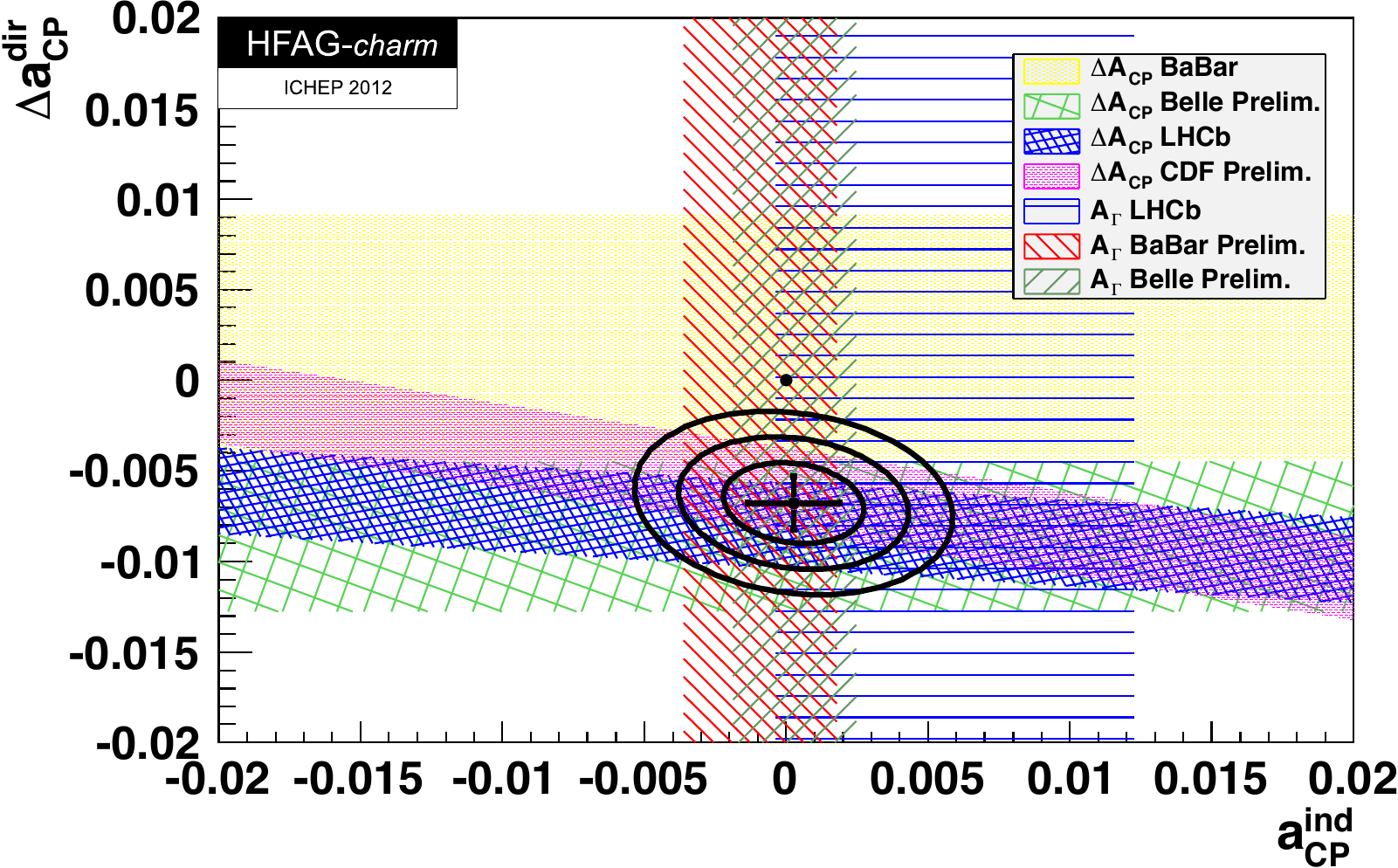}
\caption{Combined measurements of $\Delta A_{CP}$ and $A_\Gamma$, where the bands represent $\pm 1\sigma$ intervals. The point of no CP violation $(0,0)$ is shown as a filled circle, and two-dimensional 68\% CL, 95\% CL, and 99.7\% CL regions are plotted as ellipses with the best fit value as a cross indicating the one-dimensional uncertainties in their center. Reproduced from~\protect\cite{gersabeck.2011,hfag.site}.} 
\label{fig:hfagfit}
\end{center}
\end{figure}

Including higher order corrections, the first term in the right side of Eq.~\ref{eq:dacp}~\cite{gersabeck.2012} changes to $\Delta a^{dir}_{CP}[1+y_{CP}(\langle t(K^-K^+)\rangle+\langle t(K^-K^+)\rangle)/(2\tau_{D^0})]$. Assuming that the current precision of $A_\Gamma$ is such that it is not sensitive to $\Delta a^{dir}_{CP}$, HFAG fits the contributions of direct and indirect CPV to the most recent results of $y_{CP}$, $A_\Gamma$ and $\Delta A_{CP}$ by Babar, Belle, CDF and LHCb~\cite{hfag.site}, as shown in Fig.~\ref{fig:hfagfit}. This fit yields $\Delta a^{dir}_{CP}=(-6.78\pm 1.47)\times 10^{-3}$ and $a^{ind}_{CP}=(0.27\pm 1.63)\times 10^{-3}$, again favouring a direct CPV effect over an indirect, and is consistent with CP conservation with a confidence level of only 0.002\%. 

BaBar and Belle searched for CPV in the decays of the $D^+$ and $D^+_s$ mesons to two-body final states containing a $K^0_S$ using data samples corresponding to 469 fb$^{-1}$~\cite{babar.dk0spi,babar.dk0s} and 673 fb$^{-1}$~\cite{belle.dk0s}, respectively. The results are presented in Table~\ref{tab:dk0s}. The first two decays are SCS decays, while the latter result from the coherent sum of Cabibbo favoured and doubly Cabibbo suppressed transitions and are not expected to exhibit CPV in the charm dynamics.  For all those decays, however, the SM predicts a small CP asymmetry due to the $K^0-\bar{K}^0$ mixing. This contribution is measured to be $(+0.332\pm0.006)$\% for the $D^+_s\to K^0_S\pi^+$ mode and $(-0.332\pm0.006)$\% for the other modes~\cite{k0sinducedAcp}. The Belle result for the $D^+\to K^0_s\pi^+$ asymmetry was performed with a larger sample, corresponding to the full data set of 977 fb$^{-1}$, and deviates from zero with a significance of 3.2$\sigma$. In this work, they estimate a correction to the $K^0-\bar{K}^0$ mixing induced asymmetry following~\cite{k0sinducedAcp.corr} and, after subtracting this contribution, they estimate a CP asymmetry due to the charm dynamics alone of $(-0.024\pm 0.094\pm 0.067)$\%, which is the most precise measurement of $A_{CP}$ in charm decays to date. The $A_{CP}$ values obtained for the other channels are all consistent with zero and with the SM expectation. In particular the Babar measurements presented in May 2011 are the most precise~\cite{babar.dk0s} determination of these asymmetries. 

\begin{table}[h]
\begin{center}
\caption{Most recent experimental results for CP asymmetries in the decays $D^+\to K^0_SK^+$,  $D^+_s\to K^0_S\pi^+$, $D^+\to K^0_S\pi^+$ and $D^+_s\to K^0_SK^+$.}
\protect\label{tab:dk0s}
\small
\begin{tabular}{ccc}\hline
Decay                  & BaBar $A_{CP}$(\%)      &   Belle $A_{CP}$(\%)\\\hline
 $D^+\to K^0_SK^+$     & $+0.13\pm0.36\pm0.25$\cite{babar.dk0s}   &   $-0.16\pm0.58\pm0.25$\cite{belle.dk0s}  \\ 
 $D^+_s\to K^0_S\pi^+$ & $+0.6\pm2.0\pm0.3$\cite{babar.dk0s}      &   $+5.45\pm 2.50\pm0.33$\cite{belle.dk0s}\\
 $D^+_s\to K^0_SK^+$   & $-0.05\pm 0.23\pm 0.24$\cite{babar.dk0s} &   $+0.12\pm0.36\pm0.22$\cite{belle.dk0s}\\
 $D^+\to K^0_S\pi^+$   & $-0.44\pm 0.13\pm0.10$\cite{babar.dk0spi}  & $-0.363\pm0.094\pm0.067$\cite{belle.dk0spi} \\\hline 
\end{tabular} 
\end{center}
\end{table}

\subsection{Multi-body decays} 

In the decays of D mesons to final states containing more than 2 hadrons, the rich resonant structure of the intermediate state provide the strong phases necessary to the observation of CPV, under the hypothesis of existing weak CPV phases. The interference between the resonances can lead to CPV asymmetries which vary across the phase space. These local asymmetries can be large compared to the overall asymmetry or can even happen in the absence of phase space integrated asymmetry.

LHCb used a model independent approach to look for a CPV signal in the SCS decay $D^+\to K^-K^+\pi^+$ using a high purity sample ($\sim$90\%) containing about 370$\times 10^3$ decays selected from about 35 pb$^{-1}$ of data collected in 2010~\cite{lhcb.dkkpi}. Within this technique~\cite{bediaga.anis,babar.anis}, the Dalitz plot is divided in bins and for each bin the statistical significance of the asymmetry  
\begin{equation}
S^i_{CP}= \frac{N^i_{+}-\alpha N^i_-}{\sqrt{N^i_{+}+\alpha^2 N^i_-}}
\end{equation} 
is calculated, where $N^i_{\pm}$ is the number of $D^{\pm}$ decays in each bin $i$ and $\alpha$ is the ratio between the total number of $D^+$ and $D^-$ events and is used as a correction due to a global production asymmetry. The distribution of $S^i_{CP}$ is normal under the hypothesis of CP conservation. A $\chi^2$ test using $\chi^2=\sum^{N}_{i=1}S^i_{CP}$ provides a numerical evaluation for the degree of confidence for the assumption that the differences between the $D^+$ and $D^-$ Dalitz plots are driven only by statistical fluctuations. Four different binning schemes are considered, two of them uniform and two adaptative which account for the resonant structure of the decay. 
Higher statistics control samples consisting of the CF decays $D^+_s\to K^-K^+\pi^+$ and $D^+\to K^-\pi^+\pi^+$ are used to demonstrate that possible instrumental asymmetries varying over the Dalitz plot are under the sensitivity of the method. The method is also applied to the events in the side bands of the $D^+$ mass distribution to prove that no effect can be introduced by the background. The p-values obtained with the $\chi^2$ tests for the $S^i_{CP}$ distributions using the four different binning choices are 12.7\%, 10.6\%, 82.1\% and  60.5\%. The widths and means obtained from the fits to the distributions are all consistent with 1 and 0. No evidence for CPV is found in any of the binning schemes. 


BaBar has also performed a similar analysis of $D^+\to K^-K^+\pi^+$ events~\cite{babar.milind}, using a sample of 476 fb$^{-1}$ of data with $\sim228\times 10^3$ events (92\% pure). They divide the Dalitz plane into 100 equally populated bins and find a 72\% probability for the consistency with the assumption of no CPV. The measured integrated asymmetry is $A_{CP}=(0.35 \pm 0.30 \pm 0.15)$\%. A full amplitude analysis provides CP-violating differences for the phases and magnitudes of the intermediate resonances, all consistent with zero.

CDF searched for CPV in the resonant substructure of $D^{*+}$ tagged $D^0\to K^0_S\pi^-\pi^+$ decays using a sample corresponding to 6 fb$^{-1}$ collected at Run II~\cite{cdf.dkspipi}. The SM predictions for this channel are of the order of 10$^{-6}$, dominated by the $K^0_S$ mixing contribution. The selected sample is about 90\% pure and contains approximately $350\times 10^3$ events. A simultaneous maximum likelihood fit to the $D^0$ and $\bar{D}^0$ Dalitz plots, taking into account the effect of efficiency over the Dalitz space, provide resonances parameters and CP-violating fractions, amplitudes and phases. Differences between the $p_T$ distributions of the slow pions of opposite charges are used to reweight the $\bar{D}^0$ distribution. All the asymmetries are measured to be consistent with zero, as is the overall integrated asymmetry, $A_{CP} = (−0.05\pm 0.57\pm 0.54)$\%. 
A model independent analysis similar to the one described above for the $D^+\to K^-K^+\pi^+$ decay also provides null CPV signal. 

For the decay $D^0\to \pi^-\pi^+\pi^-\pi^+$, the phase space becomes 5-dimensional. LHCb performed the first model independent CPV analysis of this decay, using about $180\times 10^3$ $D^{*+}$ tagged candidates, selected from a sample of 1 fb$^{-1}$ collected in 2011~\cite{lhcb.d24pi}. The decay $D^0\to K^-\pi^+\pi^-\pi^+$ is used as control mode and events in the $D^0$ sample are rejected in a way such to ensure that the $\eta$ and $p_T$ distributions of the $D^0$ decays match those of the $\bar{D}^0$ decays. The distribution of the asymmetry significances across the phase space is shown to be compatible with the hypothesis of CP conservation.  


\section{Summary and Conclusions}

Thanks for the pioneering work of the charm factories and the excellent work done by Babar, Belle and CDF,
 mixing in the $D^0-\bar{D}^0$ system is well established, though no single measurement with a significance above 5 standard deviations was available until the beginning of the preparation of these proceedings. In November 2012, LHCb has submitted a paper reporting the observation of oscillations in the time dependent rate of $D^0\to K^+\pi^-$ relative to $D^0 \to K^-\pi^+$, which excludes the no mixing hypothesis with a significance of 9.1 $\sigma$~\cite{lhcb.mixingobserved}. 

The recent results obtained for $\Delta A_{CP}$ raised the interest about CPV in charm decays. A lot of effort was put, from the theory side, to understand on whether this can be, if confirmed, due to a new physics effect or to an enhancement of SM penguin contributions~\cite{cpv.np.sm}. Until now, no clear picture emerged yet. More precise measurements of the difference of asymmetries and of the individual asymmetries $A_{CP}(K^-K^+)$ and $A_{CP}(\pi^-\pi^+)$ are needed. Advances in the theoretical understanding of charm decays can also help to disentangle this puzzle, as well as precise measurements in additional neutral and charged decay modes. CDF, BaBar and Belle ended operation and are finishing to analyse their samples. LHCb has shown the ability to reconstruct charm decays with purities comparable to the experiments at $e^+e^-$ machines and has already collected the largest charm sample of the world, corresponding to a luminosity of 1 fb$^{-1}$ at $\sqrt{s}=7$ TeV and 2 fb$^{-1}$ at $\sqrt{s}=8$ TeV. Precisions comparable or below the SM predictions will only be achievable at the LHCb upgrade~\cite{lhcb.upgrade} and the super B factories~\cite{bfactories1,bfactories2}.

\section*{Acknowledgements}I am grateful to the organizers of PIC2012 and to my colleagues at the LHCb charm working group for useful discussions. My participation at this conference was supported by CNPq and FINEP.

\end{document}